\documentclass[journal]{IEEEtran}

\usepackage[utf8]{inputenc} 
\usepackage[T1]{fontenc} 
\usepackage[cmex10]{amsmath} 

\usepackage{amssymb} 
\usepackage{mathtools} 
\usepackage{amsfonts} 

\usepackage[dvips]{graphicx} 
\graphicspath{{./}}
\DeclareGraphicsExtensions{.eps}

\usepackage{xcolor} 

\usepackage{lineno}

\usepackage{bbm} 
\usepackage{xfrac} 
\usepackage{cancel} 
\usepackage{wrapfig} 
\usepackage{tensor} 

\usepackage{theorem}
\usepackage{cite}

\usepackage{url} 

\usepackage{psfrag} 

\usepackage[inline]{enumitem} 

\usepackage{array} 

\usepackage{tabu} 
\newcolumntype{M}[1]{>{\centering\arraybackslash}m{#1}}
\newcolumntype{R}[1]{>{\arraybackslash}m{#1}}
\newcolumntype{N}{@{}m{0pt}@{}}

\usepackage{aliascnt} 

\usepackage[hidelinks,linktocpage]{hyperref}

\usepackage{tikz}
\usetikzlibrary{positioning,arrows,decorations.pathmorphing,shapes}

\usepackage{xparse} 

\usepackage{ifthen} 

\newcommand{\mynewtheorem}[2]{
  \newaliascnt{#1}{dummy}
  \newtheorem{#1}[#1]{#2}
  \aliascntresetthe{#1}
  \expandafter\def\csname #1autorefname\endcsname{#2}
}

\theoremstyle{plain}
\theorembodyfont{\normalfont}

  \mynewtheorem{theorem}{Theorem$\!$}
  \mynewtheorem{lemma}{Lemma$\!$}
  \mynewtheorem{corollary}{Corollary$\!$}

  \mynewtheorem{definition}{Definition$\!$}

  \mynewtheorem{xmpl}{Example$\!$}
  \newenvironment{example}{\begin{xmpl}\hspace*{-1ex}}{\hfill$\Box$\end{xmpl}}

\newtheorem{cnstr}{Construction$\!$}

\newenvironment{construction}{\begin{cnstr}}{\hfill$\Box$\end{cnstr}}

\newtheorem{regime}{Regime$\!$}

\newcommand\numberthis{\stepcounter{equation}\tag{\theequation}}

\newcounter{enumrom}
\renewcommand{\theenumrom}{(\roman{enumrom})}

\makeatletter
\renewcommand{\@endtheorem}{\endtrivlist}
\makeatother

\makeatletter
\renewcommand{\thefigure}{{\@arabic\c@figure}}
\renewcommand{\fnum@figure}{{\bf Figure\,\thefigure}}
\makeatother

\renewcommand{\leq}{\leqslant}

\renewcommand{\geq}{\geqslant}

\newcommand{\cC}{\mathcal{C}}

\newcommand{\cH}{\mathcal{H}}

\newcommand{\cM}{\mathcal{M}}

\newcommand{\cP}{\mathcal{P}}

\newcommand{\cS}{\mathcal{S}}
\newcommand{\cT}{\mathcal{T}}

\renewcommand{\Bbb}{\mathbb}

\newcommand{\N}{{\Bbb N}}
\newcommand{\R}{{\Bbb R}}

\newcommand{\Z}{{\Bbb Z}}

\DeclarePairedDelimiter\abs{\lvert}{\rvert}
\DeclarePairedDelimiter\ceilenv{\lceil}{\rceil}
\DeclarePairedDelimiter\floorenv{\lfloor}{\rfloor}
\DeclarePairedDelimiter\parenv{\lparen}{\rparen}
\DeclarePairedDelimiter\sparenv{\lbrack}{\rbrack}
\DeclarePairedDelimiter\bracenv{\lbrace}{\rbrace}
\DeclarePairedDelimiterX\mathset[2]{\lbrace}{\rbrace}{#1 \mathrel{}\delimsize\vert\mathrel{} #2}
\DeclarePairedDelimiterX\inner[2]{\langle}{\rangle}{#1 \mathrel{},\mathrel{} #2}

\DeclareDocumentCommand\norm{ o m }{
    \IfNoValueTF{#1}
        {\left\Vert#2\right\Vert}
        {\left\Vert#2\right\Vert_{#1}}
}

\DeclareDocumentCommand\der{ o m o }{
    \IfNoValueTF{#1}
        {
            \IfNoValueTF{#3}
                {\frac{d}{d{#2}}}
                {\frac{d{#3}}{d{#2}}}
        }
        {\parenv*{\frac{d}{d{#2}}}^{#1}\IfNoValueTF{#3}{}{#3}}
}
\DeclareDocumentCommand\partder{ o m m }{
    \IfNoValueTF{#1}
        {\frac{\partial{#3}}{\partial{#2}}}
        {\frac{\partial^{#1}{#3}}{{\partial{#2}}^{#1}}}
}
\DeclareDocumentCommand\df{ o m o }{
    d\IfNoValueTF{#1}{}{^{#1}}{#2}\IfNoValueTF{#3}{}{_{#3}}
}

\newcommand{\desc}[2]{\overset{#2}{\underset{#1}{\implies}}}

\DeclareMathOperator{\irr}{Irr}
\DeclareMathOperator{\rll}{RLL}
\DeclareMathOperator{\wt}{wt}

\DeclareMathOperator{\pref}{Pref}
\DeclareMathOperator{\suff}{Suff}
\DeclareMathOperator{\cpct}{cap}

\newcommand{\eqdef}{\triangleq}
\def\equationautorefname#1#2\null{%
  Eq.#1(#2\null)%
}

\begin{document}
\title{Reconstruction Codes for DNA Sequences with Uniform Tandem-Duplication Errors}
\author{Yonatan~Yehezkeally,~\IEEEmembership{Student~Member,~IEEE,}
        and~Moshe~Schwartz,~\IEEEmembership{Senior~Member,~IEEE}
        \thanks{This work was presented in part at {ISIT}'2018.}%
        \thanks{This work was supported in part by the Israel Science Foundation (ISF) under grant no.~270/18.}%
        \thanks{The authors are with the School of Electrical and Computer Engineering, Ben-Gurion University of the Negev, Beer Sheva 8410501, Israel
  (e-mail: yonatany@post.bgu.ac.il; schwartz@ee.bgu.ac.il).}%
        \thanks{Copyright (c) 2019 IEEE.  Personal use of this material is permitted.  Permission from IEEE must be obtained for all other uses, in any current or future media, including reprinting/republishing this material for advertising or promotional purposes, creating new collective works, for resale or redistribution to servers or lists, or reuse of any copyrighted component of this work in other works.}
}
\maketitle
%
\begin{abstract}
DNA as a data storage medium has several advantages, 
including far greater data density compared to electronic 
media. We propose that schemes for data storage in the 
DNA of living organisms may benefit from studying the 
reconstruction problem, which is applicable whenever 
multiple reads of noisy data are available. This strategy 
is uniquely suited to the medium, which inherently 
replicates stored data in multiple distinct ways, caused 
by mutations.
We consider noise introduced solely by uniform 
tandem-duplication, and utilize the relation to 
constant-weight integer codes in the Manhattan metric. 
By bounding the intersection of the cross-polytope with 
hyperplanes, we prove the existence of reconstruction 
codes with full rate, as well as suggest a construction 
for a family of reconstruction codes.
\end{abstract}

\section{Introduction}

\IEEEPARstart{D}{NA} is attracting considerable attention in 
recent years as a medium for data storage, due to its high 
density and longevity \cite{ChuGaoKos12}. Data storage in DNA 
may provide integral memory for synthetic-biology methods, 
where such is required, and offer a protected medium for 
long-period data storage \cite{Bal13,WonWonFoo03}. 
In particular, storage in the DNA of living organisms is now 
becoming feasible \cite{Shi17}; it has varied usages, 
including watermarking genetically modified organisms 
\cite{AriOha04,HeiBar07,LisDauBruKliHamLeiWag12} or research 
material \cite{WonWonFoo03,JupFicSamQinFig10}, and even affords 
some concealment to sensitive information \cite{CleRisBan99}. 
Naturally, therefore, data integrity in such media is of great 
interest.

Several recent works have studied the inherent constraints 
of storing and retrieving data from DNA. While desired 
sequences (over quaternary alphabet) may be synthesized 
(albeit, while suffering from substitution noise), 
generally data can only be read by observation of its 
subsequences, quite possibly an incomplete observation 
\cite{KiaPulMil16}. Moreover, the nature of DNA and current 
technology results in asymmetric errors which depend upon 
the dataset \cite{GabKiaMil17}. The medium itself also 
introduces other types of errors which are atypical in 
electronic storage, such as symbol/block-deletion and 
adjacent transpositions (possibly complemented) 
\cite{GabYaaMil18}. Finally, the purely combinatorial 
problem of recovering a sequence from the multiset of all 
its subsequences (including their numbers of incidence), 
was also studied, e.g., \cite{AchDasMilOrlPan15,ShoCouTse16}, 
as well as coding schemes involving only these multisets (or 
their profile vectors -- describing the incidence frequency 
of each subsequence) \cite{RavSchYaa19}.

Other works were concerned with data storage in the DNA 
of a living organism. While this affords some level of 
protection to the data, and even propagation (through DNA 
replication), it is also exposed to specific noise 
mechanisms due to mutations. Examples of such noise 
include symbol insertions, deletion, substitutions 
(point-mutation), and duplication (including tandem- and 
interspersed-duplication). 
Therefore, schemes for data storage in live DNA must 
address data integrity and error-correction.

In an effort to better understand these typical noise mechanisms,
their potential to generate the diversity observed in nature was
studied. \cite{HasSchBru16} classified the \emph{capacity} and/or
\emph{expressiveness} of the systems of sequences over a finite
alphabet generated by four distinct substring duplication rules:
end-duplication, tandem-duplication, palindromic-duplication, and
interspersed-duplication.  \cite{JaiHasBru17} fully characterized the
expressiveness of bounded tandem-duplication systems, proved bounds on
their capacity (and, in some cases, even exact values).
\cite{JaiFarSchBru17b} later showed that when point-mutations act
together with tandem-duplication as a sequence-generation process,
they may actually increase the capacity of the generated system.
\cite{AloBruHasJai17} looked at the typical \emph{duplication
  distance} of binary sequences; i.e., the number of
tandem-duplications generating a binary sequence from its root. It was
proven that for all but an exponentially small number of sequences
that number is proportional to the sequence length. Further, when
tandem-duplication is combined with point-mutations (here, only within
the duplicated string), it was shown that the frequency of
substitutions governs whether that distance becomes logarithmic.

The generative properties of interspersed-duplication 
were also studied from a probabilistic point of view. 
\cite{FarSchBru15,FarSchBru19} showed (under assumption of uniformity) 
that the frequencies of incidence for each subsequence 
converge to the same limit achieved by an i.i.d. source, 
thus reinforcing the notion that interspersed-duplication 
is--on its own--capable of generating diversity. 
\cite{EliFarSchBru16} specifically looked at tandem- and 
end-duplication, and found exact capacities in the case of 
duplication length $1$ by a generalization of the 
P{\'{o}}lya urn model that applies to strings. It also 
tightly bounded the capacity of complement 
tandem-duplication, a process where the duplicated symbol 
is complemented (using binary alphabet).

Finally, error-correcting codes for data affected by 
tandem-duplication have been studied in 
\cite{JaiFarSchBru17a}, which presented a construction of 
optimal-size codes for correcting any number of errors 
under \emph{uniform tandem-duplication} (fixed duplication 
length), computing their (and thus, the optimal-) capacity. 
It also presented a framework for the construction of 
optimal codes for the correction of a fixed number of 
errors. Next, it studies bounded tandem-duplications, where 
a characterization of the capacity of error-correcting 
codes is made for small constants. In general, it 
characterized the cases where the process of 
tandem-duplication can be traced back uniquely.
More recently, a flurry of activity in the subject includes
works such as \cite{LenJunWac18,KovTan18a,LenWacYaa19}
which provide some implicit and explicit constructions for
uniform tandem-duplication codes, as well as some bounds.

However, classical error-correction coding ignores some 
properties of the DNA storage channel; namely, stored 
information is expected to be replicated, even as it is 
mutated. This lends itself quite naturally to the 
reconstruction problem \cite{Lev01}, which assumes that 
data is simultaneously transmitted over several noisy 
channels, and a decoder must therefore estimate that data 
based on several (distinct) noisy versions of it.
Solutions to this problem have been studied in several 
contexts. It was solved in \cite{Lev01} for sequence 
reconstruction over finite alphabets, where several error 
models were considered, such as substitutions, 
transpositions and deletions. Moreover, a framework was 
presented for solving the reconstruction problem in general 
cases of interest in coding theory, utilizing a graph 
representation of the error model, which was further 
developed in \cite{LevKonKonMol08,LevSie09}. 
The problem was also studied in the context of permutation 
codes with transposition and reversal errors 
\cite{Kon07,KonLevSie07,Kon08}, and partially solved therein. 
Later, applications were found in storage technologies 
\cite{CasBla11,YaaBruSie16,YaaBru18,CheKiaVarVuYaa18}, 
since modern application might preclude the retrieval of a 
single data point, in favor of multiple-point requests. 
However, the problem hasn't been addressed yet for data 
storage in the DNA of living organisms, where it may be most 
applicable.

In this paper, we study the reconstruction problem over DNA sequences,
with uniform tandem-duplication errors. The main contributions of the
paper are the following: We show that reconstruction codes in this
setting are necessarily error-correcting codes with appropriately
chosen minimum distance, based on the uncertainty parameter. We also
show that in two asymptotic regimes, we can always obtain higher size
than error-correcting codes. These asymptotic regimes include what we 
believe is the most interesting one, where the uncertainty is sublinear, 
and the time (number of mutations) is bounded by a constant. 

The paper is organized as follows: In \autoref{sec:preliminaries} we
present notations and definitions. In \autoref{sec:codes} we
demonstrate that reconstruction codes partition into error-correcting 
codes and find the requisite minimal-distance of each part, as a 
function of the reconstruction parameters. 
We see that these parts can be isometrically embedded as 
constant-weight codes in the Manhattan metric. Finally, in 
\autoref{sec:capacity} we show that reconstruction codes exist with 
full capacity, and also suggest a construction for reconstruction 
codes; we also briefly review recent results, published after the 
submission of this paper. We conclude with closing remarks in 
\autoref{sec:conclusion}.

\section{Preliminaries}\label{sec:preliminaries}

Throughout this paper, though DNA is composed of four nucleotide 
bases, we observe the more general case of sequences over a finite 
alphabet; since the alphabet elements are immaterial to our discussion, 
we denote it throughout as $\Z_q$. We observe the set of finite sequences 
(also: \emph{words}) over it $\Z_q^* \eqdef \bigcup_{n=0}^\infty \Z_q^n$. 
For any two words $u,v\in\Z_q^*$, we denote their concatenation $uv$. For 
each word $x\in \Z_q^n$, we denote its \emph{length} $\abs*{x} = n$. We 
also take special note of the set of words with length higher than or 
equal to some $0<k\in\N$, which we denote 
$\Z_q^{\geq k} \eqdef \mathset*{x\in\Z_q^*}{\abs*{x}\geq k}$. For ease of 
notation, we let $\N$ stand for the set of non-negative integers 
throughout the paper; when an integer is assumed to be strictly positive, 
we make special note of that fact.

For $0<k\in\N$, $i\in\N$, we define a \emph{tandem-duplication} of 
\emph{duplication-length} $k$ by the mappings 
\[
\cT_{k,i}(x) \eqdef \begin{cases}
uvvw & x = uvw,\ \abs*{u}=i, \abs*{v}=k,\\
x & \text{otherwise}.
\end{cases}
\]
If $y = \cT_{k,i}(x)$ and $y\neq x$ (which occurs whenever 
$\abs*{x}\geq i+k$), we say that $y$ is a \emph{descendant} of $x$, 
and denote $x\desc{k}{}y$. In what follows, we focus on the uniform 
tandem-duplication model (i.e., we fix $k$) because of its simplicity.

Further, given a sequence $\bracenv*{x_j}_{j=0}^t\subseteq\Z_q^*$ such 
that for all $0\leq j<t$, $x_j\desc{k}{} x_{j+1}$, we say that $x_t$ is 
a \emph{$t$-descendant} of $x_0$, and denote $x_0\desc{k}{t}x_t$. For 
completeness, we also denote $x\desc{k}{0}x$. Finally, if there exists 
some $t\in\N$ such that $x\desc{k}{t}y$, we also denote $x\desc{k}{*}y$.

We denote the set of $t$-descendants of $x\in\Z_q^*$ as 
\[
D^t_k(x) \eqdef \mathset*{y\in\Z_q^*}{x\desc{k}{t} y},
\]
for some $t\in\N$.
We also denote the \emph{descendant cone} of $x$ by 
$D_k^*(x) \eqdef \bigcup_{t=0}^\infty D_k^t(x)$.

We say that $x\in\Z_q^{\geq k}$ is \emph{irreducible} if 
$x\in D_k^*(y)$ implies $y=x$.
We exclude from the definition shorter words, for which the condition 
vacuously holds.
We denote by $\irr_k$ the set of all irreducible words, and 
$\irr_k(n) \eqdef \irr_k\cap\Z_q^n$.

It was shown in \cite{LeuMarMit05,JaiFarSchBru17b} that for each word 
$x\in\Z_q^{\geq k}$, a unique irreducible word exists for which $x$ is a 
descendant. We call it the \emph{root} of $x$, and denote it by $R_k(x)$. 
This induces an equivalence relation by $x\sim_k y$ if $R_k(x) = R_k(y)$.

We also follow \cite{JaiFarSchBru17b} in defining, for 
$x\in\Z_q^{\geq k}$, $\pref_k(x)$ as the length-$k$ \emph{prefix} of 
$x$, and $\suff_k(x)$ as its suffix; 
i.e., if $x = u u' = v' v$ where $\abs*{u}=\abs*{v}=k$, then 
$\pref_k(x) = u$ and $\suff_k(x)=v$.
Using this notation, we define an embedding 
$\phi_k:\Z_q^{\geq k}\to\Z_q^k\times\Z_q^*$ by
\[
\phi_k(x) \eqdef \parenv*{\pref_k(x), 
\suff_{\abs*{x}-k}(x)-\pref_{\abs*{x}-k}(x)}.
\]
It is seen in \cite{JaiFarSchBru17b} that this mapping is indeed 
injective. Further, it was shown that, defining 
$\zeta_{k,i}:\Z_q^k\times\Z_q^*\to \Z_q^k\times\Z_q^*$ by
\[
\zeta_{k,i}(a,b) \eqdef \begin{cases}
(a,b_1 0^k b_2) & b = b_1 b_2,\ \abs*{b_1}=i,\\
(a,b) & \text{otherwise},
\end{cases}
\]
where $0<k\in\N$, $i\in\N$, we have 
\[
\phi_k\parenv*{\cT_{k,i}(x)} = \zeta_{k,i}\parenv*{\phi_k(x)}.
\]
The simplicity of $\zeta_{k,i}$ in comparison to $\cT_{k,i}$ motivates 
the analysis of tandem-duplications using the $\phi_k$ images of 
sequences.

If $b\in\Z_q^*$ is composed of the subsequences 
\[
b = 0^{s_1}w_1 0^{s_2}\cdots w_m 0^{s_{m+1}};\quad
w_1,\ldots,w_m \in \Z_q\setminus\bracenv*{0}
\]
we define
\begin{align*}
\mu(b) &\eqdef 0^{s_1 \bmod k}w_1 0^{s_2 \bmod k}\cdots 
w_m 0^{s_{m+1} \bmod k} ,\\
\sigma(b) &\eqdef \parenv*{\floorenv*{\frac{s_1}{k}},\ldots,
\floorenv*{\frac{s_{m+1}}{k}}}.
\end{align*}
We may note that $\wt_H(b) = \wt_H(\mu(b))=m$, where $\wt_H$ is the Hamming weight, and 
$\sigma(b)\in \N^{\wt_H(b)+1} = \N^{\wt_H(\mu(b))+1}$.
We also observe that $b$ is recoverable from $\sigma(b),\mu(b)$.
It was proven in \cite[Cor.~10]{JaiFarSchBru17b} that if 
$\phi_k(x) = (a,b)$ then
\[
\phi_k\parenv*{R_k(x)} = (a,\mu(b)).
\]
Thus, if $x,y\in\Z_q^{\geq k}$, $\phi_k(x) = (a_1,b_1)$ and 
$\phi_k(y) = (a_2,b_2)$, then $x\sim_k y$ if and only if $a_1=a_2$ and 
$\mu(b_1) = \mu(b_2)$. Moreover, $x\in\irr_k$ if and only if 
$\sigma(b_1)=(0,0,\ldots,0)$. Note that, equivalently, we may say 
that $b$ contains no zero-runs of length $k$; such sequences are 
called \emph{$(0,k-1)_q$-Run-Length-Limited}, or $(0,k-1)_q$-$\rll$.

For $x\in\irr_k$, $\phi_k(x)=(a,b)$, we denote $m(x) \eqdef \wt_H(b)$
and define $\psi_x:D_k^*(x)\to \N^{m(x)+1}$ by 
$\psi_x(y) \eqdef \sigma(b')$, where $\phi_k(y) = (a,b')$.

Finally, for $n\geq k$ and $x,y\in\Z_q^n$ we define
\[
d_k(x,y) \eqdef \min\mathset*{t\in\N}{D_k^t(x)\cap D_k^t(y)\neq\emptyset}
,\]
or $d_k(x,y) = \infty$ if 
$\mathset*{t\in\N}{D_k^t(x)\cap D_k^t(y)\neq\emptyset} = \emptyset$.
It was shown in \cite[Lem.~14]{JaiFarSchBru17b} that 
$d_k(x,y) = \infty$ if and only if $x\not\sim_k y$, hence 
$d_k(\cdot,\cdot)$ is finite on $D_k^t(x)$, for any particular 
$x\in\Z_q^{\geq k}$. Furthermore, \cite[Lem.~19]{JaiFarSchBru17b} shows 
that for any $x\sim_k y$ with $\abs*{x}=\abs*{y}$ it holds that 
\[
d_k(x,y) = \tfrac{1}{2}\norm[1]{\sigma(b_1)-\sigma(b_2)},
\]
thus $d_k(\cdot,\cdot)$ defines a metric on each equivalence
class of $\sim_k$.

\section{Reconstruction Codes}\label{sec:codes}

The reconstruction problem in the context of uniform tandem-duplication 
errors can be stated as follows: 
suppose data is encoded in $C\subseteq\Z_q^n$, and suppose we later are 
able to read distinct $x_0,x_1,\ldots,x_N\in D_k^t(c)$ for some specific 
$c\in C$ and $t\in\N$; can we uniquely identify $c$?

It is apparent (see \cite{Lev01}) that to allow successful reconstruction 
we require codes to satisfy the following.

\begin{definition}\label{ftr}
Take $N,t,n>0$. We say that $C\subseteq \Z_q^n$ is a \emph{uniform 
tandem-duplication reconstruction code}, which we abbreviate as an 
$(N,t,k)_q$-UTR code, if 
\[
\max\mathset*{\abs{D_k^t(c)\cap D_k^t(c')}}{c,c'\in C,\ c\neq c'} \leq N.
\]
\end{definition}

The purpose of this section is to characterize reconstruction 
codes. By an evaluation of the size of intersection of descendant 
cones, we determine the achievable size of $(N,t,k)_q$-UTR codes. 
We shall state the solution to this problem in terms of 
error-correcting codes for the Manhattan metric, and devote the next 
section to an observation of such codes.

\subsection{Structure of descendant cones}

Throughout this section we fix some $x\in \irr_k$, and denote 
$\phi_k(x) = (a,b)$.

As noted above, for all $y\in D_k^*(x)$, we have 
$\phi_k(y) = (a,b')$, with $\mu(b') = b$ (hence, in particular, 
$\wt_H(b') = \wt_H(b)$).
We therefore denote $m = m(x) = \wt_H(b)$ and make the following 
definition:

\begin{definition}
We let $\psi_x:D_k^*(x)\to \N^{m+1}$ be defined by 
$\psi_x(y) = \sigma(b')$, where $\phi_k(y) = (a,b')$.
\end{definition}

It was noted in the previous section that $\psi_x$ is then 
distance-preserving from $\parenv*{D_k^*(x), d_k}$ to 
$\parenv*{\N^{m+1}, \frac{1}{2}\norm[1]{\cdot}}$ 
(the definition of $d_k$, made here specifically for sequences of 
equal-length, can be extended to $D_k^*(x)$ by considering the shortest 
path between any two sequences, but for simplicity in what follows, we 
shall implicitly only consider $d_k$ as a metric over $D_k^t(x)$ for any 
given $t\in\N$).

\begin{definition}
We define on $\N^{m+1}$ the partial order $b'\preceq b''$ if for every 
coordinate $i=1,\ldots,m+1$ it holds that $b'_i\leq b''_i$ ($\preceq$ 
is the well-known product order).
\end{definition}

The poset $\parenv*{\N^{m+1}, \preceq}$ has a simple structure. We shall 
therefore find it more convenient to consider $D_k^*(x)$ in these terms:
\begin{lemma}\label{lem:posetiso}
$\psi_x$ is a poset isomorphism from $\parenv*{D_k^*(x), \desc{k}{*}}$ 
to $\parenv*{\N^{m+1}, \preceq}$.
In particular, 
\begin{enumerate}
\item\label{posetiso:en1}
For all $y,y'\in D_k^*(x)$ there exists $z\in D_k^*(y)\cap D_k^*(y')$ 
such that
\[
D_k^*(y)\cap D_k^*(y') = D_k^*(z);
\]

\item\label{posetiso:en2}
If in addition $\abs*{y} = \abs*{y'}$ then for all $t\in\N$
\[
\abs*{D_k^t(y)\cap D_k^t(y')} = \begin{cases}
0 & t < d_k(y,y'), \\
\abs*{D_k^{t-d_k(y,y')}(x)} & t \geq d_k(y,y').
\end{cases}
\]
\end{enumerate}
\end{lemma}
\begin{IEEEproof}
We note that $x\in\irr_k$, hence $\psi_x(x) = (0,0,\ldots,0)\in\N^{m+1}$. 
Further, we note that in the image of $\phi_k$, a tandem-duplication 
$\zeta_{k,i}$ corresponds to increasing by one a single coordinate of 
$\sigma(\cdot)$, i.e., an addition of a unit vector $e_j\in\N^{m+1}$ to 
$\psi_x(\cdot)$.

Hence, $\psi_x$ is indeed a poset isomorphism, and we see that 
$\desc{k}{*}$ endows $D_k^*(x)$ with a lattice structure; We denote 
the \emph{join} of $y,y'\in D_k^*(x)$ as $y\vee y'$, and their 
\emph{meet} $y\wedge y'$. It follows that $z = y\vee y'$ satisfies 
\autoref{posetiso:en1}.

Finally, if $\abs*{y} = \abs*{y'}$ then by definition of $d\eqdef 
d_k(y,y')$ we have $z = y\vee y' \in D_k^d(y)\cap D_k^d(y')$, and 
\autoref{posetiso:en2} is now straightforward to prove from the 
poset-isomorphism.
\end{IEEEproof}

Given \autoref{lem:posetiso}, we can now find the size of intersection 
of descendant cones for any $c,c'\in\Z_q^n$ ($n\geq k$), keeping in mind 
that $D_k^*(c)\cap D_k^*(c')\neq \emptyset$ if and only if $c\sim_k c'$.

\begin{lemma}\label{lem:cone-cut-size}
For $x\in\irr_k$, $\abs*{D_k^t(x)} = \binom{t+m(x)}{m(x)}$.
\end{lemma}
\begin{IEEEproof}
By \autoref{lem:posetiso} we know that 
\[
D_k^t(x) = \mathset*{y\in D_k^*(x)}{\norm[1]{\psi_x(y)} = t}.
\]
Since $\psi_x:D_k^*(x)\to\N^{m(x)+1}$ is bijective, $\abs*{D_k^t(x)}$ 
equals the number of distinct integer solutions to 
$\sum_{j=1}^{m+1} x_j = t$, where $x_1,\ldots,x_{m+1}\geq 0$ 
(equivalently, the number of distinct ways to distribute $t$ identical 
balls into $m(x)+1$ bins).
\end{IEEEproof}

\subsection{Size of reconstruction codes}

In this section we aim to estimate the maximal size of $(N,t,k)_q$-UTR 
codes.

\begin{definition}
For $m,r>0$ we denote the \emph{simplex} of \emph{dimension} $m$ and 
\emph{weight} $r$, or $(m,r)-$simplex 
\[
\Delta^m_r \eqdef \mathset*{\parenv*{x_i}_{i=1}^{m+1}\in\N^{m+1}}
{\sum_{j=1}^{m+1} x_j = r}.
\]
\end{definition}

\begin{theorem}\label{th:size-eq}
We take positive integers $N,t$ and $n > k$. For $C\subseteq \Z_q^n$ and 
$x\in\irr_k$ we partition $C_x \eqdef C\cap D_k^*(x)$ and define 
$r(x) \eqdef \frac{n-\abs*{x}}{k}$.

If $C_x\neq\emptyset$ then $r(x)\in\N$ and 
$r(x) < \floorenv*{\frac{n}{k}}$. Moreover, $C$ is an $(N,t,k)_q$-UTR 
code if and only if for all $x\in\irr_k$ such that $C_x\neq\emptyset$, 
the image $\psi_x(C_x) \subseteq \Delta^{m(x)}_{r(x)}$ satisfies 
\[
\min\mathset*{\tfrac{1}{2}\norm[1]{c-c'}}{c\neq c'\in \psi_x(C_x)} 
\geq d_{N,t}\parenv*{m(x)},
\]
where we make the notation 
\[
d_{N,t}(m) \eqdef \min\mathset*{\delta\in\N}{\binom{t-\delta+m}{m}\leq N}.
\]
\end{theorem}
\begin{IEEEproof}
If $C\cap D_k^*(x)\neq\emptyset$ then it follows from the definitions
that for some $r\in\N$ we have $\abs*{x}+rk=n$; since $\abs*{x}\geq
k$, necessarily $r=r(x) < \floorenv*{\frac{n}{k}}$. Furthermore,
${C\cap D_k^*(x)} = {C\cap D_k^r(x)}$, hence we have seen in the proof
of \autoref{lem:posetiso} that for all $y\in D_k^r(x)$ we have $\psi_x(y)
= \sum_{u=1}^r e_{j_u} \in \Delta^{m(x)}_r$.

In addition, by \autoref{lem:posetiso} and \autoref{lem:cone-cut-size}, 
for all $x\in\irr_k$ and $y\neq y'\in C_x$ the size of intersection 
$D_k^t(y)\cap D_k^t(y')$ is $\binom{t-d_k(y,y')+m(x)}{m(x)}$. It follows 
that $C_x$ is an $(N,t,k)_q$-UTR code if and only if that size 
is no greater than $N$ for all such $y,y'\in C_x$.

Recalling that $\psi_x$ is bijective and distance-preserving, i.e., that 
$d_k(y,y') = \frac{1}{2}\norm[1]{\psi_x(y)-\psi_x(y')}$, the claim 
follows for $C_x$.

To conclude the proof, we recall that for $x,x'\in\irr_k$ we have 
$D_k^*(x)\cap D_k^*(x')=\emptyset$, 
hence $C$ is an $(N,t,k)_q$-UTR if and only if the same is true 
for $C_x$, for all $x\in\irr_k$.
\end{IEEEproof}

In other words, \autoref{th:size-eq} states that the intersection of a 
uniform-tandem-duplication reconstruction code $C$ with the descendant 
cone of any irreducible word $D_k^*(x)$ can be viewed as an 
error-correcting code with a suitable minimal distance. 
Further, we see that these error-correcting codes are equivalent to 
codes in the Manhattan metric over a simplex $\Delta^{m(x)}_{r(x)}$.
We note here, however, that this does not hold for $C$ in general: not 
only is each code's minimal distance dependent on $x$, but the dimension 
and weight of the simplex in which that code exists do, as well.

We therefore see that constructions and bounds on the size of 
error-correcting codes for uniform tandem-duplication depend on doing 
the same for error-correcting codes in the Manhattan metric over 
$\Delta^m_r$. We start by notating the maximal size of such codes:

\begin{definition}
For $m,r>0$ and $d\geq 0$ we define 
\[
M(m,r,d) 
\eqdef \max\mathset[\Big]{\abs*{C}}{C\subseteq \Delta^m_r,\; 
\min_{\substack{\mathclap{c,c'\in C}\\ c\neq c'}} 
\tfrac{1}{2}\norm[1]{c-c'}\geq d}.
\]
\end{definition}

We now reiterate that if $C\subseteq \Z_q^n$, $x,x'\in\irr_k(n-rk)$ 
(i.e., $r(x)=r(x')=r$) and $m(x)=m(x')$, then 
$D_k^{n-rk}(x)\cong D_k^{n-rk}(x')$ (through, e.g., 
$\psi_{x'}^{-1}\circ\psi_x$). It is therefore practical 
to assume $\abs*{C_x} = \abs*{C_{x'}} = M\parenv*{m, r, d_{N,t}(m)}$ for 
all such $x,x'$. This results in the following corollary, which 
concludes this section: 

\begin{corollary}\label{cor:size-eq}
If $C\subseteq \Z_q^n$ is an $(N,t,k)_q$-UTR code, and for all 
$x\in\irr_k$ it holds that $\abs*{C_x} = M\parenv*{m, r, d_{N,t}(m)}$,
then
\begin{align*}
\abs*{C} &= \sum_{r=0}^{\floorenv*{n/k}-1} \sum_{m} 
M\parenv*{m, r, d_{N,t}(m)} \cdot \\
&\qquad\qquad\qquad\quad \cdot 
\abs*{\mathset*{x\in\irr_k(n-rk)}{m(x)=m}} \\
&= \sum_{r=0}^{\floorenv*{n/k}-1} \sum_{m} 
M\parenv*{m, r, d_{N,t}(m)} \cdot q^k \cdot \\
&\qquad\qquad\qquad\quad \cdot 
\abs*{\mathset*{b\in \Z_q^{n-(r+1)k}}{\begin{smallmatrix}
b\ \text{is}\ (0,k-1)_q\text{-}\rll \\
\wt_H(b) = m
\end{smallmatrix}}}
\end{align*}
\end{corollary}
\begin{IEEEproof}
First, trivially, $\abs*{C} = \sum_{x\in\irr_k}\abs*{C_x}$.

Observe that $x\in\irr_k$ satisfies $C_x\neq \emptyset$, $r(x)=r$ and 
$m(x)=m$, if and only if $x\in \Z_q^{n-rk}$ and in $\phi_k(x)=(a,b)$, 
$b$ is $(0,k-1)_q$-$\rll$, and $\wt_H(b) = m$.

The rest now follows from \autoref{th:size-eq}.
\end{IEEEproof}

\autoref{cor:size-eq} motivates us to estimate the optimal size of 
error-correcting codes in the Manhattan metric over the $(m,r)$-simplex. 
This topic was examined in some depth in \cite{KovTan18}, where a 
construction based on Sidon sets (of particular interest for our 
application, see \cite{KovTan17}, and references therein) was proposed, 
leading to lower bounds tighter than the Gilbert-Varshamov bound. For 
our purposes, we cite an asymptotic result (we slightly rephrase):
\begin{lemma}\cite[Eq.~36]{KovTan18}\label{th:lin-simplex-GV}
Take $\mu\in (0,1)$, $\rho>0$ and integer sequences $(m_n)_{n>0}$, 
$(r_n)_{n>0}$ such that $\lim_{n\to\infty}\frac{m_n}{n}=\mu$ and 
$\lim_{n\to\infty}\frac{r_n}{n}=\rho$. Also take a fixed $d>0$. Then 
\begin{align}
\lim_{n\to\infty}\frac{1}{n}\log_2 M\parenv*{m_n,r_n,d} 
= (\mu+\rho)H\parenv*{\frac{1}{1 + \frac{\rho}{\mu}}}.
\end{align}
\end{lemma}

\subsection{Minimal distance of reconstruction codes}

Next, 
before we can ascertain the sizes of error-correcting codes over 
simplices, we bound their requisite minimal distance. That is,
given $N,t>0$ and $m>0$, we establish bounds on 
\[
d_{N,t}(m) \eqdef \min\mathset*{\delta\in\N}{\binom{t-\delta+m}{m}\leq N}
\]
seen in \autoref{th:size-eq}.

\begin{lemma}\label{lem:mindist-triv}
If $N\leq m$ then  $d_{N,t}(m) = t$.
\end{lemma}
\begin{IEEEproof}
We may verify by substitution that $\delta=t$ satisfies 
$\binom{t-\delta+m}{m}\leq N$, while $\delta = t-1$ does not.
Using the strict monotonicity of $s\mapsto\binom{s+m}{m}$, we are done.
\end{IEEEproof}

In order to find a practical bound for $d_{N,t}(m)$ when $N>m$, we 
first require the following three lemmas:
\begin{lemma}\label{lem:binom-log}
\begin{enumerate}
\item\label{binom-log:en1}
\cite[Ch.10,~Sec.11,~Lem.7]{MacSlo78}
For integers $0<k<n$ it holds that 
\[
\sqrt{\frac{n}{8 k(n-k)}} 2^{n H\parenv*{\frac{k}{n}}}
\leq \binom{n}{k} 
\leq \sqrt{\frac{n}{2\pi k(n-k)}} 2^{n H\parenv*{\frac{k}{n}}}
\]
where $H$ is the binary entropy function, defined by
$H(p) \eqdef -p\log_2 p - (1-p)\log_2(1-p)$.

\item\label{binom-log:en2}
\[
n H\parenv*{\frac{k}{n}} - \frac{1}{2}\log_2(2n) 
\leq \log_2\binom{n}{k} < n H\parenv*{\frac{k}{n}}.
\]
\end{enumerate}
\end{lemma}
\begin{IEEEproof}
For \autoref{binom-log:en2}, we see that if  $0<k<n$ we have 
$n-1 \leq k(n-k)\leq \frac{n^2}{4}$, hence 
\begin{align*}
\frac{n}{2\pi k(n-k)} &\leq \frac{1}{2\pi}\parenv*{1+\frac{1}{n-1}} 
\leq \frac{1}{\pi} < 1, \\
\frac{n}{8k(n-k)} &\geq \frac{1}{2n}.
\end{align*}
Thus the claim trivially follows from \autoref{binom-log:en1}.
\end{IEEEproof}

For ease of notation in what follows, we make the notation, for 
$1\leq x\in \R$:
\[
\cH(x) \eqdef x H\parenv*{\frac{1}{x}}.
\]

\begin{lemma}\label{lem:min-d_cH}
For $N>m>0$ and $t>0$ it holds that 
\[
d_{N,t}(m) \leq \min\mathset*{\delta\in\N}
{\cH\parenv*{1+\frac{t-\delta}{m}}\leq \frac{\log_2 N}{m}}.
\]
\end{lemma}
\begin{IEEEproof}
Under the assumption, $\delta=t-1$ satisfies the inequality 
$\binom{t-\delta+m}{m}\leq N$. Therefore we may restrict the minimum to 
$\delta < t$, giving $0 < m < (t-\delta)+m$. Now, \autoref{lem:binom-log} 
implies 
\[
\log_2\binom{t-\delta+m}{m} \leq m\parenv*{1+\frac{t-\delta}{m}} 
H\parenv*{\frac{1}{1+\frac{t-\delta}{m}}},
\]
which completes the proof.
\end{IEEEproof}
\begin{lemma}
\label{lem:xH-sq}
For $x\geq 1$ it holds that 
$\cH(x) \leq 2\sqrt{x-1}$.
\end{lemma}
\begin{IEEEproof}
The claim 
can be restated by the substitution $p=\frac{1}{x}$ as the known 
inequality $H(p)^2\leq 4p(1-p)$ (its proof follows elementary calculus, 
and is omitted here).
\end{IEEEproof}

Finally,
\begin{theorem}\label{th:min-d_log}
Take $N > m > 0$. Then 
\[
d_{N,t}(m) \leq \max\bracenv*{1, t - \floorenv*{\frac{(\log_2 N)^2}{4m}}}.
\]
\end{theorem}
\begin{IEEEproof}
Using \autoref{lem:xH-sq} we may bound 
$\cH\parenv*{1+\frac{t-\delta}{m}} \leq 2\sqrt{\frac{t-\delta}{m}}$. 
\autoref{lem:min-d_cH} therefore implies that it suffices to require 
$2\sqrt{\frac{t-\delta}{m}} \leq \frac{\log_2 N}{m}$, and reordering 
the inequality we get $\delta \geq t-\frac{(\log_2 N)^2}{4m}$, yielding 
the claim.
\end{IEEEproof}

\section{Capacity of reconstruction codes}\label{sec:capacity}

\begin{definition}
We define the \emph{rate} of a code $C\subseteq \Z_q^n$ as 
\[
R(C)\eqdef \frac{1}{n}\log_q\abs*{C},
\]
and the capacity of a system $\cC\subseteq \Z_q^*$ as 
\[
\cpct(\cC) \eqdef \limsup_{n\to\infty} 
\frac{1}{n}\log_q\abs*{\cC\cap \Z_q^n}.
\]
\end{definition}

We are interested in $\sup\bracenv*{\cpct(\cC)}$, where $\cC$ is any 
family of reconstruction codes (i.e., $\cC\cap\Z_q^n$ is an 
$(N_n,t_n,k)_q$-code for all $n$).

The purpose of this section is to determine that optimal capacity in 
two asymptotic regimes:
\begin{regime}\label{reg:time-bounded}
When $N_n = o(n)$ and $t_n=t$ is fixed.
\end{regime}
\begin{regime}\label{reg:time-linear}
When $N_n = 2^{\alpha n}$ and $t_n = \beta n$ for constants 
$\alpha,\beta > 0$ (such that $N_n,t_n\in\N$ for some, hence infinitely 
many, indices).
\end{regime}
In practical applications, \autoref{reg:time-bounded} is likely to 
apply, since we may indeed expect the number of duplications $t$, which 
is dependent on the period of time before data is read, to be fixed 
w.r.t. $n$. The allowed uncertainty $N_n$ will also likely be bounded. 
\autoref{reg:time-linear} requires \autoref{th:min-d_log} (and some 
restrictions over the values of $\alpha,\beta$), but allows us to 
calculate capacity in much the same way, which we do after presenting 
the first.

Note, since \cite{JaiFarSchBru17a} showed that $\irr_k(n)$ can correct 
any number of tandem-duplication errors, they are trivially 
$(N,t,k)_q$-codes for all $N,t$ (more precisely, they are 
$(0,t,k)_q$-codes for all $t$). 
In comparison, in the setting we consider only $t$ tandem-duplications 
are assumed to have occurred, therefore the codes we seek are less 
restrictive. Nevertheless, at the time of this paper's submission no 
bounds on the size of error-correcting codes for a fixed number of 
tandem-duplications were known;
It is our purpose, then, to demonstrate that reconstruction codes exist 
which have strictly higher capacity than $\irr_k$, and suggest 
constructions for families of such codes.

First, we denote for any $n,r\in\N$ such that $n\geq k$ and 
$r<\floorenv*{\frac{n}{k}}$, and any $N,t\in\N$ 
\begin{align*}
&\cM_{N,t}(n,r) \eqdef
\sum_{m} M\parenv*{m, r, d_{N,t}(m)} \cdot \\
&\qquad\cdot \abs*{\mathset*{b\in \Z_q^{n-(r+1)k}}{\begin{smallmatrix}
b\ \text{is}\ (0,k-1)_q\text{-}\rll \\
\wt_H(b) = m
\end{smallmatrix}}}.
\end{align*}
We recall for all $n$, if $r_n = \arg\max_r \cM_{N,t}(n,r)$, that
by \autoref{cor:size-eq} we have an $(N,t,k)_q$-code
$C\subseteq\Z_q^n$ with $\abs*{C} \geq q^k \cM_{N,t}(n,r_n)$. 
\autoref{cor:size-eq} also implies that for all $C\subseteq\Z_q^n$ it 
holds that $\abs*{C} \leq \frac{n}{k} q^k \cM_{N,t}(n,r_n)$. We 
therefore focus on maximizing
$\limsup_{n\to\infty} \frac{1}{n}\log_q \cM_{N,t}(n,r_n)$ by choice of
$r_n$.

In what follows, we take $\gamma\in(0,1)$ and set 
$r_n = \frac{1-\gamma}{k}n - 1$ for any $n\in\N$ for which $r_n\in\N$; 
we shall assume that such $n$ exist (hence, infinitely many exist), and 
refer only to such indices.

For all $x\in\irr_k(n-r_n k) = \irr_k\parenv*{k + \gamma n}$, recall 
that we denoted $\phi_k(x) = (a,b)$ with $b\in \Z_q^{\gamma n}$ in 
$(0,k-1)_q\text{-}\rll$. We shall build a reconstruction code in 
the descendant cones of only such $x$, which we denote $C_\gamma$.

\begin{lemma}\label{lem:rll-weight}
There exists a system $\cS\subseteq (0,k-1)_q\text{-}\rll$ and 
$\theta\in \parenv*{\frac{1}{2},1}$ such that 
\[
\cpct (\cS) 
= \lim_{l\to\infty} \frac{1}{l}\log_q \abs*{\cS\cap \Z_q^l} 
= \cpct\parenv*{(0,k-1)_q\text{-}\rll}
\]
and for all $b\in \cS$ it holds that $\wt_H(b) \geq \theta\abs*{b}$.
\end{lemma}
\begin{IEEEproof}
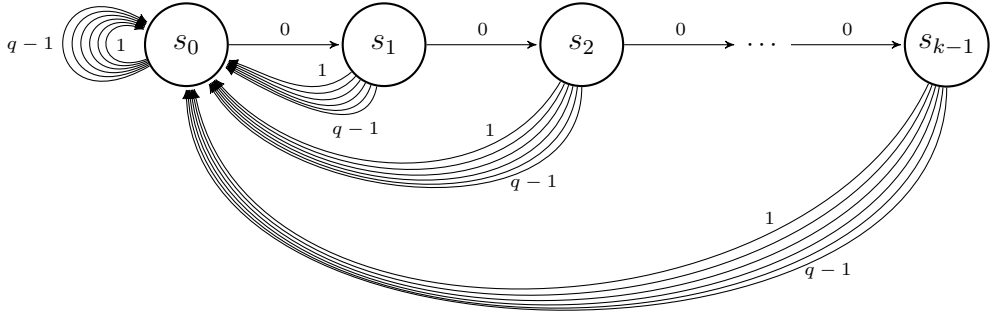
\begin{figure*}
\begin{center}
\begin{tikzpicture}[->,>=stealth',shorten >=1pt,auto,node distance=1.5cm,
    main node/.style={thick,circle,draw,font=\sffamily\large,minimum size=1.1cm}]

  \node[main node] (0) {$s_0$};
  \node[main node] (1) [right=of 0]{$s_1$};
  \node[main node] (2) [right=of 1]{$s_2$};
  \node[auto=false] (4) [right=of 2] {$\cdots$};
  \node[main node] (3) [right=of 4]{$s_{k-1}$};

  \path[every node/.style={above, midway, font=\scriptsize}]
    (0) edge node{$0$} (1)
    (1) edge node{$0$} (2)
    (2) edge node{$0$} (4)
    (4) edge node{$0$} (3);

  \path[every node/.style={midway, font=\scriptsize}]
    (0) edge[looseness=5, out=200, in=160] node(a)[right]{$1$} (0)
    (0) edge[looseness=5.6, out=202, in=158] (0)
    (0) edge[looseness=6.2, out=204, in=156] (0)
    (0) edge[looseness=6.8, out=206, in=154] (0)
    (0) edge[looseness=7.4, out=208, in=152] (0)
    (0) edge[looseness=8, out=210, in=150] node(b)[left]{$q-1$} (0);

  \path[every node/.style={near start, font=\scriptsize}]
    (1) edge[out=220, in=340] node(c)[above]{$1$} (0)
    (1) edge[out=228, in=338] (0)
    (1) edge[out=236, in=336] (0)
    (1) edge[out=244, in=334] (0)
    (1) edge[out=252, in=332] (0)
    (1) edge[out=260, in=330] node(d)[below]{$q-1$} (0);

  \path[every node/.style={near start, font=\scriptsize}]
    (2) edge[out=240, in=310] node(e)[above]{$1$} (0)
    (2) edge[out=246, in=308] (0)
    (2) edge[out=252, in=306] (0)
    (2) edge[out=258, in=304] (0)
    (2) edge[out=264, in=302] (0)
    (2) edge[out=270, in=300] node(f)[below]{$q-1$} (0);

  \path[every node/.style={near start, font=\scriptsize}]
    (3) edge[out=245, in=280] node(g)[above]{$1$} (0)
    (3) edge[out=250, in=278] (0)
    (3) edge[out=255, in=276] (0)
    (3) edge[out=260, in=274] (0)
    (3) edge[out=265, in=272] (0)
    (3) edge[out=270, in=270] node(h)[below]{$q-1$} (0);

\end{tikzpicture}\vspace*{-30pt}
\caption{The graph $G_q(k-1)$ generating the $(0,k-1)_q\text{-}\rll$ 
system.\label{fig:RLL}}
\end{center}
\end{figure*}
Let $G_q(k-1)$ be the strongly connected deterministic digraph
representing the $(0,k-1)_q\text{-}\rll$ system, seen in 
\autoref{fig:RLL}, whose adjacency matrix is
\[
T_q(k-1) = \begin{pmatrix}
q-1 & 1 & 0 & \cdots & 0 \\
q-1 & 0 & 1 &  & \vdots \\
\vdots & \vdots &  & \ddots & 0 \\
q-1 & 0 & \cdots & 0 & 1 \\
q-1 & 0 & \cdots & \cdots & 0
\end{pmatrix}
\]
As is well known for the case of 
$q=2$ (see, e.g., \cite{ZehWol88,How89}), its characteristic polynomial 
is
\[
p^{(k-1)}_q(x) 
= x^k - (q-1)\sum_{j=0}^{k-1} x^j = \frac{x^{k+1} - q x^k + (q-1)}{x-1},
\]
hence the Perron eigenvalue $\lambda$ of $T_q(k-1)$ is the unique 
positive root of $\hat{p}^{(k-1)}_q(x) = x^{k+1} - q x^k + (q-1)$ greater 
than $1$ (in fact, $\lambda\in (q-1,q)$, which can readily be confirmed 
either using elementary calculus or by information-theoretic methods, 
since $\parenv*{\Z_q\setminus\bracenv*{0}}^* \subseteq 
(0,k-1)_q\text{-}\rll \subseteq \Z_q^*$).

Further, $T_q(k-1)$ has positive right- and left-eigenvectors associated 
with $\lambda$, which we denote  $\bar{v}, \bar{w}$ respectively; 
specifically, 
\begin{align*}
\bar{v} 
&= \left(1, \lambda - (q-1), \ldots, 
\lambda^{j-1} - (q-1)\sum_{i=0}^{j-2}\lambda^i,\ldots, \right.\\
&\left.\qquad\qquad\qquad\qquad\qquad\qquad \ldots
\lambda^{k-1} - (q-1)\sum_{i=0}^{k-2}\lambda^i\right), \\
\bar{w} &= \parenv*{\lambda^{k-1}, \lambda^{k-2}, \ldots, 
\lambda^{k-j}, \ldots, 1}.
\end{align*}
and we may verify that 
\begin{align*}
v_k &= \lambda^{k-1} - (q-1)\sum_{i=0}^{k-2}\lambda^i 
= \frac{1}{\lambda}\sparenv*{\lambda^k - (q-1)\sum_{j=1}^{k-1}\lambda^j} \\
&= \frac{q-1}{\lambda} > 0
\end{align*}
and $v_j = \frac{v_{j+1}+(q-1)}{\lambda}$, hence every entry of $\bar{v}$ 
is indeed positive.

Denoting $q_{i,j} = \parenv*{T_q(k-1)}_{i,j}\cdot \frac{v_j}{\lambda v_i}$, 
it follows (see, e.g., \cite{MarRotSie01}[Sec.~3.5]) that 
$Q = (q_{i,j})_{1\leq i,j\leq k}$ is stochastic, and represents a 
transition matrix of a stationary Markov chain $\cP$ on $G_q(k-1)$ 
(a probability measure on its edges set $E_q(k-1)$) satisfying 
$H(\cP) = \log_q \lambda = \cpct\parenv*{(0,k-1)_q\text{-}\rll}$.
Further, the stationary distribution of the Markov chain, i.e., 
a positive $\bar{\pi} = \parenv*{\pi_1,\ldots,\pi_k}$ such that 
$\sum_{j=1}^k \pi_j=1$ and $\bar{\pi}^T Q = \bar{\pi}^T$, is given 
by $\pi_j = \frac{\hat{\pi}_j}{\sum_{i=1}^k \hat{\pi}_i}$, where 
$\hat{\pi}$ is defined by $\hat{\pi}_j = w_j v_j$. It holds for all $j$ 
that $\pi_j$ is the sum of probabilities $\sum\cP(e)$ of edges 
terminating at the $j$'th node.

Note, then, that
\begin{align*}
\sum_{i=1}^k \hat{\pi}_i 
&= \lambda^{k-1} + \sum_{i=2}^k \sparenv*{\lambda^{k-1} 
- (q-1) \frac{\lambda^{k-1} - \lambda^{k-i}}{\lambda-1}} \\
&= \lambda^{k-1} \sparenv*{1 + (k-1)\parenv*{1-\frac{q-1}{\lambda-1}}} 
+ \frac{q-1}{\lambda-1}\sum_{i=2}^k \lambda^{k-i} \\
&= \lambda^{k-1} \sparenv*{k - (k-1) \frac{q-1}{\lambda-1}} 
+ \frac{q-1}{\lambda-1}\sum_{j=0}^{k-2} \lambda^j \\
&= \lambda^{k-1} \sparenv*{k - (k-1) \frac{q-1}{\lambda-1}} 
+ \frac{\lambda^k - (q-1)\lambda^{k-1}}{\lambda-1} \\
&= \frac{\lambda^{k-1}}{\lambda-1} \sparenv*{\lambda - k (q-\lambda)}
\end{align*}
and in particular 
$\pi_1 = \frac{\lambda-1}{\lambda - k (q-\lambda)}$. (Incidentally, it 
follows from $\pi_1\in (0,1)$ that $1 < k (q-\lambda) < \lambda$, that 
is, $q - \frac{q}{k+1} < \lambda < q - \frac{1}{k}$.)

Next, recall that for a given $\epsilon>0$, a \emph{$(\cP,\epsilon)$-%
strongly-typical} path in $G$ is a path $\gamma = \parenv*{e_1, e_2, 
\ldots, e_l}$ (denoted by its edges 
$\bracenv*{e_1,e_2,\ldots,e_l}\subseteq E_q(k-1)$) such that each 
$e\in E_q(k-1)$ appears in the path $l\cdot\tau$ times, for some $\tau$ 
satisfying $\abs*{\tau-\cP(e)}\leq \epsilon$. If we let 
$\cS_\epsilon\subseteq\Z_q^*$ be the system induced by 
$(\cP,\frac{\epsilon}{k(q-1)})$-strongly-typical paths, then it is well 
known that $\cpct (\cS_\epsilon) = \cpct\parenv*{(0,k-1)_q\text{-}\rll}$.
Note, for $b\in \cS_\epsilon$ of length $\abs*{b}=l$, which is 
generated by the path $\gamma = \parenv*{e_1,\ldots,e_l}$, $\wt_H(b)$ is 
precisely the number of edges which terminate at the first node; since 
$\gamma$ is $(\cP,\frac{\epsilon}{k(q-1)})$-strongly-typical, 
\[
\wt_H(b) 
\geq \sum_{\mathclap{\substack{e\ \text{terminates}\\ 
\text{at first node}}}} l \cdot \parenv*{\cP(e) - \frac{\epsilon}{k(q-1)}} 
= l (\pi_1 - \epsilon)
\]
To conclude the proof, note 
\begin{align*}
\lambda +& k (q-\lambda) = q + (k-1)(q-\lambda) > q \geq 2 \\
&\implies \lambda > 2 - k (q-\lambda) \\
&\implies 2(\lambda-1) > \lambda - k (q-\lambda) 
\implies \pi_1 > \frac{1}{2}
\end{align*}
Hence we can take any $0<\epsilon<\pi_1-\frac{1}{2}$, and observe that 
$\cS = \cS_\epsilon$, $\theta = \pi_1-\epsilon$ satisfy the proposition.
\end{IEEEproof}

\autoref{lem:rll-weight} implies that there exists a subset 
$S_k\subseteq \irr_k$ such that $\cpct(S_k) = \cpct(\irr_k)$, and for 
every $x\in S_k$ of length $\abs*{x} = k+\gamma n$ we have 
$m(x) \geq \ceilenv*{\theta \cdot \gamma n}$. For the rest of this 
section we only build codes $C_\gamma^n$ in the descendant cones 
of roots in $S_k$.
Note, then, that if we denote $m_n = \ceilenv*{\theta \cdot \gamma n}$ 
and $\cC_\gamma\eqdef \bigcup C^n_\gamma$, then 
\begin{align*}
\cpct(\cC_\gamma) &\geq \limsup_{n\to\infty}\frac{1}{n}\log_q 
\Big[\abs*{\irr_k(k+\gamma n)} \cdot \\
&\qquad\qquad\qquad\quad \cdot 
M\parenv*{m_n, r_n, d_{N,t}(m_n)} \Big]\\
&= \gamma \cpct\parenv*{\irr_k} + \\
&\quad
+\limsup_{n\to\infty}\frac{1}{n}\log_q M\parenv*{m_n, r_n, d_{N,t}(m_n)} 
\numberthis \label{eq:rate}
\end{align*}
We evaluate the second addend in the following theorem:

\begin{theorem}\label{th:uncertainty}
As before, we denote $r_n = \frac{1-\gamma}{k}n - 1$ and 
$m_n = \ceilenv*{\theta\cdot \gamma n}$. Then 
\begin{align*}
&\lim_{n\to\infty} \frac{1}{n} \log_q M(m_n,r_n,d_{N_n,t_n}(m_n)) = \\
&\qquad\qquad\qquad\qquad= \frac{\theta \gamma}{\log_2 q} 
\cdot \cH\parenv*{1 + \frac{1-\gamma}{k \theta \gamma}}
\end{align*}
in both of the aforementioned two regimes:
\begin{enumerate}
\item
\autoref{reg:time-bounded}: when $N_n = o(n)$ and $t_n=t$ is fixed.

\item
\autoref{reg:time-linear}: when $N_n = 2^{\alpha n}$ and $t_n = \beta n$, 
if we additionally require $\frac{\alpha^2}{\beta} > 4 \theta \gamma$.
\end{enumerate}
\end{theorem}
\begin{IEEEproof}
\begin{enumerate}\setlength{\itemsep}{2ex}
\item
Note, for sufficiently large $n$, that $N_n < \theta\cdot \gamma n 
\leq m_n$, resulting by \autoref{lem:mindist-triv} in 
$d_{N_n,t}(m_n) = t$. 
We note that $\lim_{n\to\infty}\frac{r_n}{n} = \frac{1-\gamma}{k}$ and 
$\lim_{n\to\infty}\frac{m_n}{n} = \theta \gamma$, hence by 
\autoref{th:lin-simplex-GV} 
the claim is proven when $t$ is fixed.

\item
By 
\autoref{th:min-d_log}: 
\begin{align*}
d_{N_n,t_n}\parenv*{m_n} 
&\leq \max\bracenv*{1, \beta n 
- \floorenv*{n\frac{\alpha^2 n}{4 \ceilenv*{\theta\cdot \gamma n}}}} \\
&= \max\bracenv*{1, \ceilenv*{\parenv*{\beta 
- \frac{\alpha^2 n}{4 \ceilenv*{\theta\cdot \gamma n}}}n}}.
\end{align*}

If $\frac{\alpha^2}{\beta} > 4 \theta \gamma$ then for sufficiently large 
$n$ we have $\beta < \frac{\alpha^2 n}{4 \ceilenv*{\theta\cdot \gamma n}}$, 
hence $d_{N_n,t_n}(m_n)=1$. Since it is fixed, we may now apply the same 
argument used in the previous part.
\end{enumerate}
\end{IEEEproof}

Going forward, we shall view the lower bound to $\cpct(\cC_\gamma)$,
\[
R(\gamma)\eqdef \gamma \cpct\parenv*{\irr_k} 
+ \frac{\theta \gamma}{\log_2 q} 
\cdot \cH\parenv*{1 + \frac{1-\gamma}{k \theta \gamma}},
\]
as a function of $\gamma$. Before moving on to show that it may be made 
to exceed $\cpct(\irr_k)$ by a careful choice of $\gamma$, we look at 
the following example.
\begin{example}
Set $q=k=2$. Then the Perron eigenvalue of $T_2(1)$ is 
$\lambda = \frac{1+\sqrt{5}}{2}$, and
\[
\cpct(\irr_2) = \log_2(\lambda) = \log_2\parenv*{\frac{1+\sqrt{5}}{2}} 
\approx 0.6942.
\]
In addition, any $\theta$ which is less than 
$\pi_1 = \frac{1}{2} \parenv*{1+\frac{1}{\sqrt{5}}}\approx 0.7236$ 
satisfies 
\autoref{lem:rll-weight}.

Alternatively, we may set $q=4$ (for the special case of DNA) and 
duplication-length $k=2$. Now the Perron eigenvalue of $T_4(1)$ is given 
by $\lambda = \frac{3+\sqrt{21}}{2}$, hence 
\[
\cpct(\irr_2) = \log_4(\lambda) = \log_4\parenv*{\frac{3+\sqrt{21}}{2}} 
\approx 0.9613.
\]
Further, we may choose any $\theta$ which is less than $\pi_1 = 
\frac{1}{2}\parenv*{1+\sqrt{\frac{3}{7}}} \approx 0.8273$.

$R(\gamma)$ is shown for both cases in \autoref{fig:rate}, under the 
assumptions of asymptotic regime made in \autoref{th:uncertainty}. The 
figure demonstrates that the capacity of reconstruction codes (bounded 
from below by the maximum of the curve) is greater than $\cpct(\irr_k)$.
\begin{figure}[t]
\psfrag{rrr}{$R$}
\psfrag{gamma}{$\gamma$}
\psfrag{xax}{(a)}
\psfrag{xbx}{(b)}
\begin{center}
\includegraphics[width=0.9\columnwidth]{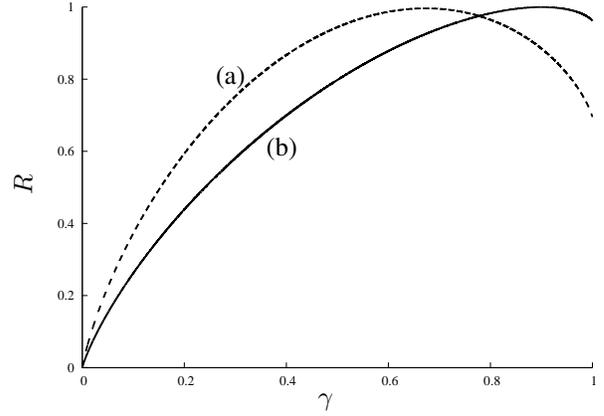}
\end{center}
\caption{Rate $R(\gamma)$ in the cases (a) $q=k=2$, $\theta=0.7236$, 
and (b) $q=4$, $k=2$, $\theta=0.8273$. 
The value at $\gamma=1$ equals $\cpct(\irr_k)$.}
\label{fig:rate}
\end{figure}
\end{example}

We now attempt to maximize $R(\gamma)$ by a proper choice of 
$\gamma\in (0,1)$.
Analysis of $R(\gamma)$ is simpler using the following change of variable:
\begin{definition}
Define $x:(0,1)\to (0,\infty)$ by 
$x(\gamma) \eqdef \frac{1-\gamma}{\gamma}$.
\end{definition}
We observe that $x(\gamma)$ is a decreasing diffeomorphism, and 
$\gamma = \frac{1}{1+x(\gamma)}$.

\begin{lemma} One has 
\begin{align*}
& R(\gamma) = \gamma\cpct(\irr_k) + \theta \gamma \left[
\parenv*{1 + \frac{x(\gamma)}{k\theta}} 
\log_q\parenv*{1+\frac{x(\gamma)}{k\theta}} \right. \\
&\qquad\qquad\qquad\qquad\qquad\qquad\qquad\qquad
\left. - \frac{x(\gamma)}{k\theta} 
\log_q\parenv*{\frac{x(\gamma)}{k\theta}}\right]
\end{align*}
\end{lemma}
\begin{IEEEproof}
We 
observe that for all $x>0$, $\log\parenv*{1+\frac{1}{x}} 
= \log\parenv*{\frac{x+1}{x}} = \log(x+1) - \log x$; in particular
\[
\log_q\parenv*{1+\frac{k\theta\gamma}{1-\gamma}} 
= \log_q\parenv*{1+\frac{1-\gamma}{k\theta\gamma}} 
- \log_q\parenv*{\frac{1-\gamma}{k\theta\gamma}}
\]
Hence,  
\begin{align*}
R(\gamma) =& \gamma\cpct(\irr_k) 
+ \frac{\theta\gamma}{\log_2 q} 
\cdot \cH\parenv*{1 + \frac{1-\gamma}{k\theta \gamma}} \\
=& \gamma\cpct(\irr_k) 
+ \theta \gamma \log_q\parenv*{1+\frac{1-\gamma}{k\theta\gamma}} \\
&\qquad\qquad\qquad\qquad
+ \frac{1-\gamma}{k}\log_q\parenv*{1+\frac{k\theta\gamma}{1-\gamma}} \\
=& \gamma\cpct(\irr_k) 
+ \parenv*{\theta \gamma + \frac{1-\gamma}{k}} 
\log_q\parenv*{1+\frac{1-\gamma}{k\theta\gamma}} \\
&\qquad\qquad\qquad\qquad\qquad\quad
- \frac{1-\gamma}{k}\log_q\parenv*{\frac{1-\gamma}{k\theta\gamma}} \\
=& \gamma\cpct(\irr_k) + \theta \gamma \bigg[
\parenv*{1 + \frac{1-\gamma}{k\theta\gamma}} 
\log_q\parenv*{1+\frac{1-\gamma}{k\theta\gamma}} \\
&\qquad\qquad\qquad\qquad\qquad\quad
- \frac{1-\gamma}{k\theta\gamma} 
\log_q\parenv*{\frac{1-\gamma}{k\theta\gamma}}\bigg]
\end{align*}
\end{IEEEproof}

We can now show that there always exists a choice of $\gamma$ for which 
we get $R(C_\gamma^n) > \cpct(\irr_k)$:
\begin{theorem}
$\max_{\gamma\in(0,1)}R(\gamma) > \cpct(\irr_k)$.
\end{theorem}
\begin{IEEEproof}
Observe that $R(\gamma)$ is continuously differentiable and satisfies 
$\lim_{\gamma\to 0}R(\gamma) = 0$, 
$\lim_{\gamma\to 1}R(\gamma) = \cpct(\irr_k)$. 
We find $R'(\gamma)$ in \autoref{fig:differential};
\begin{figure*}\scriptsize
\begin{align*}
R'(\gamma) =& \cpct(\irr_k) + \frac{d x}{d\gamma}\cdot \frac{d}{dx} 
\sparenv*{
\frac{\theta}{1+x} \parenv*{
\parenv*{1+\frac{x}{k\theta}}\log_q\parenv*{1+\frac{x}{k\theta}} 
- \frac{x}{k\theta}\log_q\parenv*{\frac{x}{k\theta}}
}}_{x=x(\gamma)} \\
=& \cpct(\irr_k) - \frac{1}{\gamma^2} \sparenv*{
\frac{-\theta}{(1+x)^2} \parenv*{
\parenv*{1+\frac{x}{k\theta}}\log_q\parenv*{1+\frac{x}{k\theta}} 
- \frac{x}{k\theta}\log_q\parenv*{\frac{x}{k\theta}}
} + \frac{\theta}{(1+x)} \cdot \parenv*{
\frac{1}{k\theta}\log_q\parenv*{1+\frac{x}{k\theta}} 
- \frac{1}{k\theta}\log_q\parenv*{\frac{x}{k\theta}}
}}_{x=x(\gamma)} \\
=& \cpct(\irr_k) + \frac{1}{k} \sparenv*{
(k\theta - 1)\log_q\parenv*{1+\frac{x(\gamma)}{k\theta}}
+ \log_q\parenv*{\frac{x(\gamma)}{k\theta}}} 
\numberthis\label{fig:differential}
\end{align*}
\end{figure*}
Thus, We can show that $R'(\gamma) = 0$ if and only if 
\begin{equation}\label{eq:max}
q^{-k \cpct(\irr_k)} 
= \parenv*{1 + \frac{x(\gamma)}{k\theta}}^{k\theta - 1} 
\cdot \frac{x(\gamma)}{k\theta}
\end{equation}
This equation has a unique solution $x_0 = x(\gamma_0)$, since the RHS is 
a monotonic increasing function of $x$, vanishing at $x=0$ and unbounded 
as $x$ grows. Moreover, $0<x_0<k\theta$, since $k\theta> 1$, hence the 
RHS is greater than $1$ at $x=k\theta$. Thus $R(\gamma)$ has a unique 
local extremum in $(0,1)$.

It now suffices to show that $R(\gamma)$ is concave, hence the 
extremum is a maximum. Indeed, 
\begin{align*}
R''(\gamma) &= \frac{1}{k} \frac{dx}{d\gamma}\cdot \frac{d}{dx}
\bigg[
(k\theta - 1)\log_q\parenv*{1+\frac{x}{k\theta}} \\
&\qquad\qquad\qquad\qquad\qquad\qquad
+ \log_q\parenv*{\frac{x}{k\theta}}
\bigg]_{x=x(\gamma)} \\
&= \frac{-1}{k\ln(q) \gamma^2} \sparenv*{
\frac{k\theta - 1}{k\theta+x(\gamma)} 
+ \frac{1}{x(\gamma)}
} < 0
\end{align*}
It follows that $R(\gamma_0)>\lim_{\gamma\to 1}R(\gamma)=\cpct(\irr_k)$.
\end{IEEEproof}

Thus, the main result of this paper is established. In what remains of 
this section we show that we can bound $\gamma_0$ which maximizes $R(\gamma)$, in practice, to any desired level of accuracy. 
We begin by establishing bounds in the following lemma.

\begin{lemma}\label{lem:init-bound-max}
Let $\gamma_0\in (0,1)$ be the unique maximum of $R(\gamma)$, and denote 
$x_0 = x(\gamma_0)$. Then
\[
x_0 \geq \frac{k\theta}{\parenv*{2^\theta q^{\cpct(\irr_k)}}^k - 1}
\]
and
\begin{align*}
x_0 &\leq \frac{1}{2}\bigg[\sqrt{\parenv*{1-q^{-\cpct(\irr_k) k}}^2 
+ k\theta q^{2-\cpct(\irr_k) k}} \\
&\qquad\qquad\qquad\qquad\qquad - \parenv*{1-q^{-\cpct(\irr_k) k}}\bigg] \\
&\leq \frac{k\theta q^2}{4\parenv*{q^{\cpct(\irr_k) k}-1}}.
\end{align*}
\end{lemma}
\begin{IEEEproof}
For fixed $x\in [0,\infty)$ define $g_x:(0,\infty)\to\R$ by 
$g_x(y) = y\ln\parenv*{1+\frac{x}{y}}$. Then
\begin{align*}
g_x'(y) 
&= \ln\parenv*{1+\frac{x}{y}} 
+ \frac{y}{1+\frac{x}{y}}\cdot \frac{-x}{y^2} 
= \ln\parenv*{1+\frac{x}{y}} - \frac{x}{y+x} \\
&= -\ln\parenv*{1-\frac{x}{x+y}} - \frac{x}{y+x} \\
&\geq -\parenv*{-\frac{x}{x+y}} - \frac{x}{y+x} \geq 0.
\end{align*}
Therefore, $f_x(y) = e^{g_x(y)} = \parenv*{1+\frac{x}{y}}^y$ satisfies 
$1+x = f_x(1) \leq f_x(y) = \parenv*{1+\frac{x}{y}}^y$ for all $y\geq 1$. 
In our case $k\theta> 1$ and $x_0$ satisfies \autoref{eq:max}, hence
\[
q^{-\cpct(\irr_k) k} 
= \parenv*{1 + \frac{x_0}{k\theta}}^{k\theta-1} \frac{x_0}{k\theta} 
\geq \frac{1+x_0}{1 + \frac{x_0}{k\theta}} \cdot \frac{x_0}{k\theta} 
= \frac{x_0 + x_0^2}{k\theta + x_0}
\]
which we simplify to 
$0 \geq x_0^2 + \parenv*{1-q^{-\cpct(\irr_k) k}} x_0 - k\theta q^{-\cpct(\irr_k) k}$. Thus, 
the first upper bound is proven. For the second, we require only that for 
$a,b>0$ it holds that $\sqrt{a+b^2}-b\leq\frac{a}{2b}$, which is readily 
shown by differentiation.

On the other hand, \autoref{eq:max} implies that $x_0\leq k\theta$. 
Therefore 
\begin{align*}
q^{-\cpct(\irr_k) k} 
= \parenv*{1 + \frac{x_0}{k\theta}}^{k\theta-1} \frac{x_0}{k\theta} 
&\leq \frac{2^{k\theta}}{1 + \frac{x_0}{k\theta}} \cdot \frac{x_0}{k\theta} \\
\iff 
k\theta q^{-\cpct(\irr_k) k} 
&\leq \parenv*{2^{k\theta} - q^{-\cpct(\irr_k) k}} x_0
\end{align*}
which proves the lower bound.
\end{IEEEproof}

Next, we show that we may tighten the bounds we derived in the previous 
lemma.

\begin{lemma}\label{lem:iter-bound-max}
Let $x_0 > 0$ be the unique solution to \autoref{eq:max}, and denote 
$z_0 = \frac{x_0}{k\theta}$. If $\underline{z} \leq z_0 \leq \overline{z}$ 
then 
$
F(\underline{z}) \leq z_0 \leq F(\overline{z})
$, 
where 
\[
F(z) \eqdef \frac{q^{-\cpct(\irr_k) k}}{\parenv*{1 
+ \frac{q^{-\cpct(\irr_k) k}}{\parenv*{1 + z}^{k\theta-1}}}^{k\theta-1}}.
\]
\end{lemma}
\begin{IEEEproof}
By assumption we have $q^{-\cpct(\irr_k) k} = \parenv*{1 + z_0}^{k\theta-1} 
\cdot z_0$, hence 
$q^{-\cpct(\irr_k) k} \leq \parenv*{1 + \overline{z}}^{k\theta-1} 
\cdot z_0$, implying that $z_0 \geq G(\overline{z})$ where 
$G(z) = \frac{q^{-\cpct(\irr_k) k}}{\parenv*{1 + z}^{k\theta-1}}$. 
Similarly, $z_0 \leq G(\underline{z})$.
The proposition now trivially follows for $F(z) = G(G(z))$.
\end{IEEEproof}

Finally, we can show that $x_0$ may be found by the following limiting 
process:
\begin{theorem}\label{th:fixed-point}
The unique solution to \autoref{eq:max} is given by 
$x_0 = k\theta\lim_{n\to\infty} F^n(z_1)$, for all $z_1\in[0,1]$.
\end{theorem}
\begin{IEEEproof}
As before, we denote the unique solution $x_0>0$, and take 
$z_0 = \frac{x_0}{k\theta}$.

Note that \autoref{lem:iter-bound-max} implies that $z_0 = F(z_0)$. We 
will prove that $F:[0,1]\to[0,1]$ is a contraction; that is, for all 
$z_1,z_2\in[0,1]$ we have $\abs*{F(z_1)-F(z_2)}\leq c\abs*{z_1-z_2}$ for 
some $c<1$. Indeed, recalling $k\theta>1$ we find 
\begin{align*}
F'(z) &= \frac{2^{-2\cpct(\irr_k) k} (k\theta- 1)^2}{(1+z)^{k\theta}
\parenv*{1 + \frac{q^{-\cpct(\irr_k) k}}{(1+z)^{k\theta-1}}}^{k\theta}} \\
&\leq \frac{(k\theta-1)^2}{(2^{2\cpct(\irr_k)})^k} 
\leq \frac{(k - 1)^2}{2^k} \leq \frac{9}{16} < 1,
\end{align*}
where the next to last inequality may be directly verified for all small 
$k$.

Having done so, we utilize Banach's fixed-point theorem to deduce that $F$ 
has a unique fixed point (necessarily $z_0$), and for all $z_1\in [0,1]$, 
defining $z_{n+1} = F(z_n)$ we get $\lim_{n\to\infty} z_n = z_0$.
\end{IEEEproof}

We can now suggest a construction for $(N,t,k)_q$-UTR codes achieving 
better capacity than the error-correcting codes $\irr_k(n)$ suggested 
in \cite{JaiFarSchBru17a} (provided that one is willing to consider 
reconstruction codes over unambiguous decoding of any single output).
\begin{construction}\label{con:a}
We set the alphabet size $q$, duplication length $k$. In the case that our 
application falls within \autoref{reg:time-bounded}, we also set a fixed 
decoding-delay $t$, and restrict the ambiguity $N_n$ to be sub-linear in 
$n$. (with the necessary adjustments, this construction also applies for 
\autoref{reg:time-linear}.)
\begin{itemize}
\item
Start by finding the Perron eigenvalue $\lambda$ of $T_q(k-1)$, and 
$\pi_1 = \frac{\lambda-1}{\lambda - k (q-\lambda)}$, as in the proof of 
\autoref{lem:rll-weight}. Set some $\theta<\pi_1$.

\item
The upper and lower bounds on $x_0$ from \autoref{lem:init-bound-max} 
can be made tighter by a repetitive application of $F(\cdot)$ from 
\autoref{lem:iter-bound-max}; \autoref{th:fixed-point} guarantees that 
the bounds--hence the acceptable error--can be made as tight as desired 
for our application.

\item
With $\gamma_0 = \frac{1}{1+x_0}$ we may find 
$r_n = \frac{1-\gamma_0}{k}n-1$, and we note that a capacity-achieving 
subset of $\irr_k(n-r_n k) = \irr_k\parenv*{k + \gamma n}$ has weight 
$m(x) \geq m_n = \ceilenv*{\theta \cdot \gamma n}$.

\item
Within $D_k^{r_n}(x)$ of just such irreducible sequences $x$ we may 
utilize any construction of codes for the Manhattan metric over 
$\Delta^{m_n}_{r_n}$ with minimal distance $t$, if it produces codes 
of size sufficiently close to $M(m_n,r_n,t)$.
For practical applications, \cite[Sec.~IV-A]{KovTan18} showed that if 
$m_n$ is a prime power, then by \cite{BosCho62} there exist such codes 
of size $\abs*{\Delta^{m_n}_{r_n}}\big/ \frac{m_n^t-1}{m_n-1}$ (which 
improves on the Gilbert-Varshamov bound, and is sufficiently tight to 
achieve the same result as in \autoref{th:uncertainty}).
\end{itemize}
\end{construction}

Note that we do not establish that \autoref{con:a} produces a system
of codes of capacity $1$, rather only greater than $\cpct(\irr_k)$.
To conclude this section, we also present a non-constructive argument
proving the existence of a system of reconstruction codes with
capacity $1$ by an application of the Gilbert-Varshamov bound.

Recall that in the proof of \autoref{th:uncertainty} we have shown that 
the minimal distance, $d_{N_n,t_n}(m_n)$ was bounded. In particular, in 
the case of interest \autoref{reg:time-bounded}, we used the fact that 
$m_n = \Theta(n)$; This does not, in general, hold for $m(R_k(y))$ for 
all $y\in\Z_q^n$.

However, if we show that to be the case for a sufficiently large subset 
$S^n\subseteq \Z_q^n$, then we may note the following: by 
\cite[Lem.~1]{KovTan18} the size of ball in the $d_k(\cdot,\cdot)$ 
metric of radius $d$ in the descendant cone of $x\in\irr_k$, where 
$m(x)\geq d$, is 
\begin{align*}
\sum_{j=0}^d & \binom{m(x)}{j} \binom{d}{j} \binom{d+m(x)-j}{d} \\
&\leq (d+1)\cdot \binom{m(x)}{d} \binom{d}{\floorenv*{d/2}} \binom{d+m(x)}{d} \\
&= O(m(x)^d) = O(n^d)
\end{align*}
It would follow that a code of size $\frac{\abs*{S^n}}{O(n^d)}$ exists 
(and, again, the capacity of these codes will be $\cpct(S^n)$).

It now suffices to show that except for a vanishingly small portion of 
$y\in\Z_q^n$, it holds that $m(R_k(y)) = \Theta(n)$. 
Indeed, recall that $m(R_k(y)) = \wt_H(\mu(b)) = \wt_H(b)$, where 
$\phi_k(y) = (a,b)$, $b\in\Z_q^{n-k}$. Then, for any real $0<\xi<1-\frac{1}{q}$,
\[
\frac{\abs{\mathset*{b\in\Z_q^{n-k}}{\wt_H(b)\leq \xi(n-k)}}}{q^{n-k}}
\leq q^{(n-k)(H_q(\xi)-1)},
\]
where $H_q(\cdot)$ is the $q$-ary entropy function,
\[ H_q(\xi) \eqdef -\xi\log_q\xi - (1-\xi)\log_q(1-\xi)+\xi\log_q(q-1),\]
and where we used a standard bounding of the size on the Hamming ball,
e.g., see \cite[Lemma 4.7]{Rot06}.

\subsection{Comparison to recent results}

Before we finish, we note here that the last argument also shows via the 
GV bound that error-correcting codes for a fixed number of 
tandem-duplications achieve capacity $1$. 
Indeed, after the submission of this manuscript 
\cite{KovTan18a,LenJunWac18} 
were made available, wherein bounds on the optimal size of such 
error-correcting codes were presented; these bounds show that the 
redundancy required to correct a fixed number of tandem-duplications 
is logarithmic in $n$.

More specifically, both works showed (see \cite[Thm.~4]{KovTan18a},
\cite[Lem.~6]{LenJunWac18}) that there exist codes $C^n\subseteq \Z_q^n$ 
that correct up to $t$ tandem-duplications, for a fixed $t\in\N$, 
satisfying 
\[
\frac{q^n}{n^t}\parenv*{\frac{q}{q-1}}^t \lesssim \abs*{C^n}
\]
(where we say that $a_n\lesssim b_n$ if $\limsup\frac{a_n}{b_n}\leq 1$). 
They also showed that the optimal size was 
$\Theta\parenv*{\frac{q^n}{n^t}}$. Finally, \cite[Lem.~3]{KovTan18a} 
demonstrated that $C^n$ can be assumed w.l.o.g. to only contain sequences 
which roots satisfy $m(x)=\Theta(n)$.

We note that error-correcting codes for $t$ tandem-duplications have 
minimal $d_k(\cdot,\cdot)$ distance $t+1$; In comparison, then, we have 
showed that $(N,t,k)_q$-UTR codes, where $t$ is fixed and $N=o(n)$, have 
minimal distance $t$ (when restricted to descendant cones of irreducible 
words with $m(x)=\Theta(n)$). 
The observations above imply that codes designed in the aforementioned 
works for correcting $t-1$ tandem-duplications, of size 
$\gtrsim \frac{q^n}{n^{t-1}}\parenv*{\frac{q}{q-1}}^{t-1}$, are 
$(N,t,k)_q$-UTR codes. 
Importantly, this validates the hypothesis that reconstruction codes for 
data storage in the DNA of living organisms offer greater data-density 
than error-correcting codes.
Namely, in comparison to the $t \log(n) + O(1)$ redundancy achieved by 
optimal error-correcting codes in \cite{KovTan18a,LenJunWac18}, 
$(N,t,k)_q$-UTR codes achieve redundancy $(t-1) \log(n) + O(1)$.

Finally, we also note for completeness that our results in 
\autoref{reg:time-linear}, albeit less applicable in practice, are unique 
to this work.

\section{Conclusion}\label{sec:conclusion}

We have proposed that reconstruction codes can be applied to data-storage 
in the DNA of living organisms, due to the channel's inherent property of 
data replication.

We have showed, under the assumption of uniform tandem-duplication noise, 
that any reconstruction code is partitioned into error-correcting codes 
for the Manhattan metric over a simplex, with minimal distances 
dependent on the reconstruction parameters. We then proved the existence 
of reconstruction codes with rate $1$, and suggested a construction of a 
family of codes, which relies on constructions of codes for the simplex. 
Via \autoref{th:fixed-point}, we showed that we can bound the parameters 
required for code-design in any real application, to any degree of 
accuracy.

We believe that further research should examine explicit code 
constructions on the simplex; specifically, encoding and decoding 
algorithms for sufficiently large codes haven't yet been developed; in 
addition, only specific asymptotic regimes have been explored, and a gap 
still exists between lower an upper bounds on the size of non-linear codes. 
It is also desirable to examine the problem under broader noise models, 
such as bounded tandem-duplication,interspersed-duplication (perhaps 
complemented), as well as combinations of multiple error models.

\section*{Acknowledgments}

The authors gratefully acknowledge the two anonymous reviewers and 
associate editor, whose insight and suggestions helped shape 
this paper.

\bibliographystyle{IEEEtranS}

%
\newpage
\begin{IEEEbiographynophoto}{Yonatan Yehezkeally}
(S'12)
is a graduate student at the School of Electrical and Computer 
Engineering, Ben-Gurion University of the Negev, Beer-Sheva, Israel. 
His research interests include coding for DNA storage, combinatorial 
structures, algebraic coding and finite group theory.

Yonatan received the B.Sc.~(\emph{magna cum laude}) and 
M.Sc.~(\emph{summa cum laude}) degrees from Ben-Gurion University 
of the Negev in 2014 and 2017 respectively, from the department of 
Mathematics and the department of Electrical and Computer Engineering.
\end{IEEEbiographynophoto}

\begin{IEEEbiographynophoto}{Moshe Schwartz}
(M'03--SM'10)
is an associate professor at the School of Electrical and Computer
Engineering, Ben-Gurion University of the Negev, Israel. His research
interests include algebraic coding, combinatorial structures, and
digital sequences.

Prof.~Schwartz received the B.A.~(\emph{summa cum laude}), M.Sc., and
Ph.D.~degrees from the Technion -- Israel Institute of Technology,
Haifa, Israel, in 1997, 1998, and 2004 respectively, all from the
Computer Science Department. He was a Fulbright post-doctoral
researcher in the Department of Electrical and Computer Engineering,
University of California San Diego, and a post-doctoral researcher in
the Department of Electrical Engineering, California Institute of
Technology. While on sabbatical 2012--2014, he was a visiting scientist
at the Massachusetts Institute of Technology (MIT).

Prof.~Schwartz received the 2009 IEEE Communications Society Best Paper
Award in Signal Processing and Coding for Data Storage.
\end{IEEEbiographynophoto}
\end{document}